\documentclass{JHEP3} 
\usepackage{epsfig,multicol}

\newcommand\fverb{\setbox\pippobox=\hbox\bgroup\verb}
\newcommand\fverbdo{\egroup\medskip\noindent%
			\fbox{\unhbox\pippobox}\ }
\newcommand\fverbit{\egroup\item[\fbox{\unhbox\pippobox}]}
\newbox\pippobox

\title{On the Potentials of Supersymmetric Theories
with Gauge-Field Mixing Terms}

\author{\speaker{Cristine Nunes Ferreira}\thanks{thanks  CNPq for financial suport.}\\
        Departamento de F\'isica, Universidade Federal do Rio de Janeiro, \\
Caixa Postal 68528, 21945-910, Rio de Janeiro, RJ, Brazil\\
        E-mail: \email{crisnfer@if.ufrj.br} }

\author{Helder  Ch\'avez\\
       Universidade Estadual do Norte Fluminense, Av. Alberto Lamego 2000,\\
Parque California, Campos dos Goytacazes, \\
28013-600, RJ, Brazil\\
       E-mail: \email{helder68@cbpf.br}}

\author{Jos\'e Abdalla Helay\"el-Neto\thanks{thanks  CNPq for financial suport.}\\
       Centro Brasileiro de Pesquisas Físicas, \\
Rua Dr. Xavier Sigaud 150, Urca,\\
22290-180, Rio de Janeiro, RJ, Brazil \\
       E-mail: \email{helayel@cbpf.br}}

\abstract{In this letter, we reconsider the delicate issue of symmetry and supersymmetry
breakings for gauge theories with gauge-field mixings. The purpose is to study generalyzed potentials in the presence of more than a single gauge potential.
In this work, following a stream of investigation on supersymmetric gauge theories without flat directions, we contemplate the possibility of building up D- and F-term potentials by means of a gauge-field mixing in connection with a $U(1)\times U(1)'$ -symmetry. We investigate a generalized potential including  an N=1 supersymmetric extension of the Maxwell-Chern-Simons model focusing on the study of cosmic string configurations. This analysis sheds some light on the formation of cosmic strings for model with violation of Lorentz symmetry. }

\begin{document}

\section{Introduction}

The motivation to consider an extra U(1)' symmetry  in gauge theories comes from  the superstring  approach \cite{1},  grand unified theories \cite{2} and  models of dynamical symmetry breaking \cite{3}. In the context of supersymmetry (SUSY),  the breaking of the extra U(1) symmetry is important to give us an expectation value to the singlet field of the Standard Model [4-7]. In superstring inspired models, the motivation for eletroweak and U(1)'  symmetry breakings can be driven by soft supersymmetry breaking parameters and yield a  Z' mass of the order of the eletroweak scale \cite{1}. The other motivation for extra Abelian factor is to find potentials without flat directions. Flat directions in  scalar potentials appear in SUSY theories: Abelian theories where the gauge symmetry is broken with a Fayet-Iliopoulos (FI) D-term \cite{Wess}. Some consequences of the SUSY teories with a D-term are  cosmic string formation [9-15]; another consequence is the hybrid inflation \cite{16}. In the first case, there is the possibility that the cosmic string has not been formated in these U(1) models with the flat directions \cite{17}. For this, we propose  other possibilities to build up a U(1) potential that, with the specific choice of parameters in the F-term and considering the gauge-field mixing or a Lorentz-breaking couplings \cite{18}, gives us a potential without flat-direction.  

The importance of the Lorentz breaking effects was proposed, a few years ago, in the context of a  Maxwell-Chern-Simons (MCS) gauge theory, as an additional magnetic moment interaction \cite{Kogan} for which Bogomol'nyi-type self-dual equations \cite{Lee}. In the context of  supersymmetry, it is important in the N=2 supersymmetric extension of the self-duality model
that relates the central charge, the extended model with  the existence of the topological quantum numbers \cite{Hlousek} and also appear naturaly in noncommutative framework \cite{9}. 

In this work, we consider the general action that exhibits these two aspects, the gauge-field mixing and the Maxwell-Chern-Simons effects  in an N=1 supersymmetric theory. We study the potential generated by these two contributions
and analyse the advantages of the use each terms. 

The letter is organized as folows. In Section 2, we discuss the generalized model with these
two contributions, the gauge-field mixing and Maxwell-Chern-Simons contribution. In Section 3, we build up the potential and discuss its validity. Section 4 is devoted to the discussion of the supersymmetry an gauge symmetry breakings in connection with a superconducting cosmic string configuration.

\section{The general supersymmetric model}       

In this section, we study the general  N=1 supersymmetric version of the
$U(1)\times U(1)'$ model with gauge mixing and Maxwell-Chern-Simons terms.
For the whole set up of superspace and superspace, the component-field
version of the $U(1)\times U(1)'$-gauge theory and  
the algebraic manipulations with the
Grassmann-valued spinorial
coordinates and fields, we refer to the conventions adopted in the
work of Ref. \cite{Piguet}.  In our approach, we can split the Lagrangian into five pieces
\begin{equation}
{\cal L} = {\cal L}_{SF} + {\cal L}_{GFM} + {\cal L}_{MCS} + {\cal L}_{D} + {\cal L}_{F},
\end{equation}
where $ {\cal L}_{SF}$ is the part of the Lagrangian that contains the scalar field couplings, given by
\begin{equation}
{\cal L}_{SF} = \bar \Phi_i  e^{2 q Q_i {\cal V}_x} \Phi_i|_{\theta \theta \bar \theta \bar \theta} +
\bar \Sigma_i  e^{2 q Q_i {\cal V}_y} \Sigma_i|_{\theta \theta \bar \theta \bar \theta} \, \, \, ,
\end{equation}
The ${\cal L}_{GFM} $ contains  the gauge-field mixing and is given by
\begin{equation}
{\cal L}_{GFM} = \alpha_1 {\cal X}^\alpha  {\cal X}_\alpha |_{\theta \theta}
+ \alpha_2{\cal Y}^\alpha{\cal Y}_\alpha |_{\theta \theta}+  \alpha_3{\cal X}^\alpha{\cal Y}_\alpha |_{\theta \theta} + h.c. \, \, \, ,
\end{equation} 
the Maxwell-Chern-Simons sector ${\cal L}_{MCS}$ reads as 
\begin{equation}
{\cal L}_{MCS} = \beta_1{\cal X}^a(D_a {\cal V}_x)S |_{\theta \bar \theta} + 
\beta_2{\cal Y}^a(D_a {\cal V}_y)S |_{\theta \bar \theta}+\beta_3{\cal X}^a(D_a {\cal V}_y)S|_{\theta \bar \theta} + h.c. \, \, 
\, ,\label{MCS}
\end{equation}
the Fayet-Iliopoullos term piece, the D-term, is
\begin{equation}
{\cal L}_{D}=   k_1 D_1 + k_2  D_2, \label{lag1}
\end{equation}
and the superpotential part, $ {\cal L}_{F}$, is proposed as below 
\begin{equation}
{\cal L}_{F}=  m\Phi_+\Phi_- + \tilde m \Sigma_+ \Sigma_-.  \label{lag1}
\end{equation}
The $\alpha$'s, $\beta $'s, $k_1$, $k_2$, m and $\tilde m$ are real parameters of the model.
The ingredient superfields of the model are  chiral scalars supermultiplets,
$\Phi $ and $\Sigma $, the gauge  superpotentials, ${\cal V}_x$, ${\cal V}_y$ 
and the dimensionless field ${\cal S}$; the latter carries the background fields responsible for the Lorentz symmetry violation. 
The $\theta $-component expansions, where  ${\cal V}_x$ and ${\cal V}_y$ are already assumed to be in the Wess-Zumino gauge, take the forms as below
\begin{equation}
\Phi = e^{-i\theta \sigma^{\mu} \bar \theta \partial_{\mu}}[\phi(x) +
\theta^{a}\xi_a(x) + \theta \theta G(x)]\, , \label{sup1}
\end{equation}
\begin{equation}
\Sigma_I(x, \theta) = \sigma_I(y) + \sqrt{2} \theta^{\alpha} \chi_{_{I}\alpha}(y) +
\theta \theta H_I(y) \, ,
\end{equation}
\begin{equation}
{\cal V}_x = \theta \sigma^{\mu} \, \bar \theta \, H_{\mu}(x) + \theta \theta \, \bar \theta \, \bar{\lambda} (x) + \, \bar \theta \bar \theta  \theta \, \lambda (x)+ \theta \theta \bar \theta \bar \theta 
D_1(x) \, , \label{sup2}
\end{equation}
\begin{equation}
{\cal V}_y = \theta \sigma^{\mu} \, \bar \theta A_{\mu}(x) + \theta \theta \bar \theta \,
\bar \chi (x) + \bar \theta \bar \theta \theta \, \chi (x)+ \theta \theta \bar \theta \bar \theta
D_2 (x) \, , \label{sup4}
\end{equation}
\begin{equation}
{\cal S} = e^{-i\theta \sigma^{\mu} \bar \theta \partial_{\mu}}[S(x) +
\theta^{a}\psi_a(x) + \theta \theta F(x)]\, ,  \label{sup3}
\end{equation}
with the superfield-strengths $ {\cal X}_a $ and ${\cal Y}_a$ written as
\begin{equation}
\begin{array}{ll}
{\cal X}_a = -\frac{1}{4} \bar D^2 D^a {\cal V}_x \, ,\\
{\cal Y}_a = -\frac{1}{4} \bar D^2 D^a {\cal V}_y \, .
\end{array}
\end{equation}
Our conventions for the SUSY covariant derivatives are  given  as follows
\begin{equation}
\begin{array}{ll}
D_a = \partial_a - i \sigma^{\mu}_{a \dot a} \bar \theta^{\dot a}
\partial_{\mu} \, , &  \\
\bar D_{\dot a} =- \partial_{\dot a} + i \theta^a \sigma^{\mu}_{\dot a a}
\partial_{\mu}\, , &
\end{array}
\end{equation}
where the $\sigma^{\mu}$-matrices read as $\sigma^{\mu} \equiv ({\bf 1}; \sigma^i)$, the $\sigma^i$'s
being the Pauli matrices.
As it can be readily checked, this action is invariant under
two independent sets of Abelian gauge transformations, with superfield  parameters $\Lambda_1$ and $\Lambda_2$
\begin{equation}
\begin{array}{llll}
\phi(x) &\rightarrow & \phi'(x)= \phi(x)e^{i\xi_1 (x)}\, , &\\
\sigma(x) &\rightarrow & \sigma'(x)= \sigma(x)e^{i\xi_2 (x)}\, , &\\
H_\mu(x) &\rightarrow & H^{\prime}_\mu(x) = H_\mu (x) - \partial_\mu \xi_1(x)\, ,\\
A_\mu(x) &\rightarrow & A^{\prime}_\mu(x) = A_\mu (x) - \partial_\mu \xi_2(x)\, ,
&
\end{array}
\end{equation}
where $\xi_1$ and $\xi_2$ are real parameters that appear as the $\theta$-independent components of $\Lambda_1$ and $\Lambda_2$, respectively.

 The Lagrangian
takes now the form
\begin{equation}
{\cal L} = {\cal L}_B + {\cal L}_F + {\cal L}_Y - U \, .
\end{equation}
We shall from now on focus on the potential $U$, since our main effort is to discuss the pattern of the internal symmetry and SUSY breakdowns.

\section{ The general potential with Chern-Simons $U(1) \times U(1)'$-mixing }

In this first analysis, let us consider the main aspects of the  potential for the 
general component Lagrangian (\ref{lag1}). The potential U is given by
\begin{equation}
U = \frac{A_1}{2}D_1^2 + \frac{A_2}{2}D_2^2 +
\frac{A_3}{2} D_1 D_2 + \sum_I \bar F_I F_I + \sum_i \bar G_i G_i\, , \label{pot}
\end{equation}
with $A_1 = \alpha_1 + 8 \beta_1 (S + S^* )$, $A_2 = \alpha_2 + 8 \beta_2(S+S^*)$ and $A_3 = \alpha_3 + 8 \beta_3(S+S^*) $. The solutions to the auxiliary fields appearing in (\ref{pot}) are given below
\begin{equation}
\begin{array}{ll}
D_1 =\frac{1}{\Gamma}\left[ \frac{A_3}{4A_2}\sum_I eE_I |\sigma_I|^2-
\frac{1}{2}\sum_i q Q_i |\phi_i|^2   + \frac{A_3}{4A_2} k_2 - k_1 \right],\\
\\
D_2 = \frac{1}{\Gamma}\left[\frac{A_3}{4A_1}\sum_i qQ_i |\phi_i|^2 -
\frac{1}{2} \sum_I eE_I |\sigma_I|^2 +
\frac{A_3}{4A_1} k_1 -k_2 \right],\\
\\
\bar G_i =- \frac{\partial W}{\partial \bar \phi_i} = -m \phi_i \, , \\
\\
\bar F_I  = - \frac{\partial W}{\partial \bar \sigma_I}= -\tilde m  \sigma_I\, ,
\end{array}
\end{equation}
considering $A_1 =1 $, $A_2 =1$ and  $A_3 = 2 A, $ we have
$\Gamma = 1- A^2$. The constant A depends on the mixing parameters and the Lorentz breaking effects; it reads as  
\begin{equation}
A = \alpha + 8 \beta (S + S^*)\, .\label{A}
\end{equation}
The Fayet-Illiopoulos  D-term provides a possible way of
spontaneously breaking SUSY \cite{Wess}.

The potential (\ref{pot}), in turn, can be split as follows
\begin{equation}
U_{CS} = U_{\phi} + U_{\sigma} +U_{\phi - \sigma},\label{U1}
\end{equation}
\noindent
where $U_{\phi}$ is the self-coupling potential for the
charged scalars, $\phi_+$ and $\phi_-$
\begin{equation}
\begin{array}{lll}
U_{\phi}& =& \frac{1}{\Gamma}\Big[ \frac{q^2}{8}(|\phi_+|^2 - |\phi_-|^2)^2 + \left( \beta m^2 - \frac{q p}{2} - \frac{A q k}{2} \right) |\phi_+|^2 +\\
&&+ \left( \beta m^2 + \frac{q p}{2} + \frac{A q k}{2} \right) |\phi_-|^2 \left. + \frac{k^2}{2} + \frac{p^2}{2} + A p k  \right]\, ,
\end{array}
\label{potcs}
\end{equation}
with $k_1 = -p$ and $k_2 = k $. 
$U_{\sigma}$ is the part that contains the $\sigma$-field self-interactions with $\tilde m =0$ 
\begin{equation}
\begin{array}{lll}
U_{\sigma} & =& \frac{1}{\Gamma }\left[ \frac{e^2}{8}(|\sigma_+|^2 - |\sigma_-|^2)^2 -
\left( \frac{A e p}{2} + \frac{ e k}{2} \right) |\sigma_+|^2 + \left( \frac{A e p}{2} + \frac{e k}{2} \right) |\sigma_-|^2 \right]\, ,
\end{array}
\end{equation}
and $U_{\phi - \sigma}$ is the mixed $\phi - \sigma$ self-interaction part, given by
\begin{equation}
U_{\phi - \sigma}= -\frac{A qe}{4 \Gamma}\left(|\phi_+|^2 -
|\phi_-|^2\right)\left(|\sigma_+|^2 - |\sigma_-|^2\right)\, .
\end{equation}
In the next section, let us analyse this potential in connection with the cosmic string configuration. We also plot the region of validity of the potential. We analyse the symmetry and supersymmetry breakdowns and discuss the induced supersymmetry breakings given by
Lorentz breaking effects.

\begin{figure}
\epsfig{file=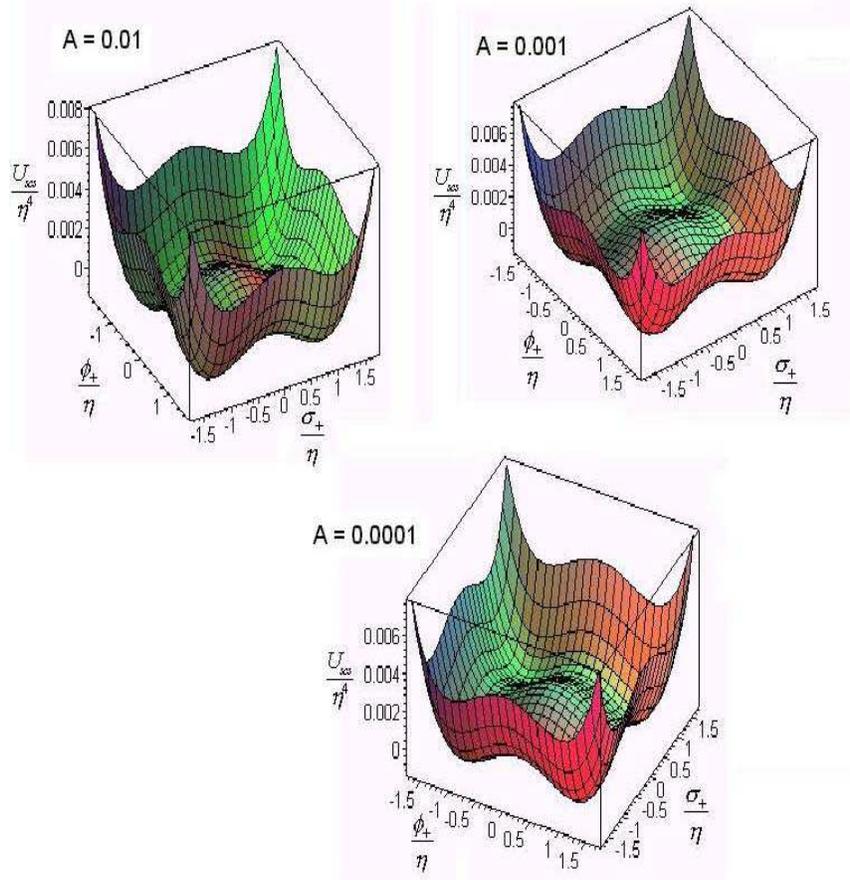, width=.9\textwidth}
\vskip -3 cm
\caption{ The potential of the superconducting cosmic string for some mixing paramenter $A$. The mixing parameters are in agreement with the cosmic string particle generation presented in \cite{11}.}
\vskip -0.1 cm
\label{fig1}
\end{figure}

\section{The application to cosmic strings}

Now, we analyse the crucial issue of gauge
symmetry and supersymmetry breakings, and the consequent formation
of a cosmic string configuration. The cosmic string that we have revised in Ref. \cite{11},  is determined in connection of the scalar potential $U_{CS}$  of the  eq.(\ref{U1}). The minimum energy configuration of a static vortex   potential for \, $U_{scs}$ \, in \, (\ref{U1})\, is \, $<\phi_+>=\eta$ , \, $<\phi_-> = 0$, \, \, \, $<\sigma_+> = 0$ and $<\sigma_->=0$ , \,
where \, $ \eta^2 =  \frac{2 p}{q}v
$ \, and \, $v = \sqrt{[1 + \frac{k^2}{p^2} + 2\frac{A k}{p}]} $, with the constraint given by 
\begin{equation}
m^2= \frac{q p}{2 \Gamma}(1  + \frac{A k}{p} - v)\, . \label{m2}
\end{equation}

\begin{figure}
\epsfig{file=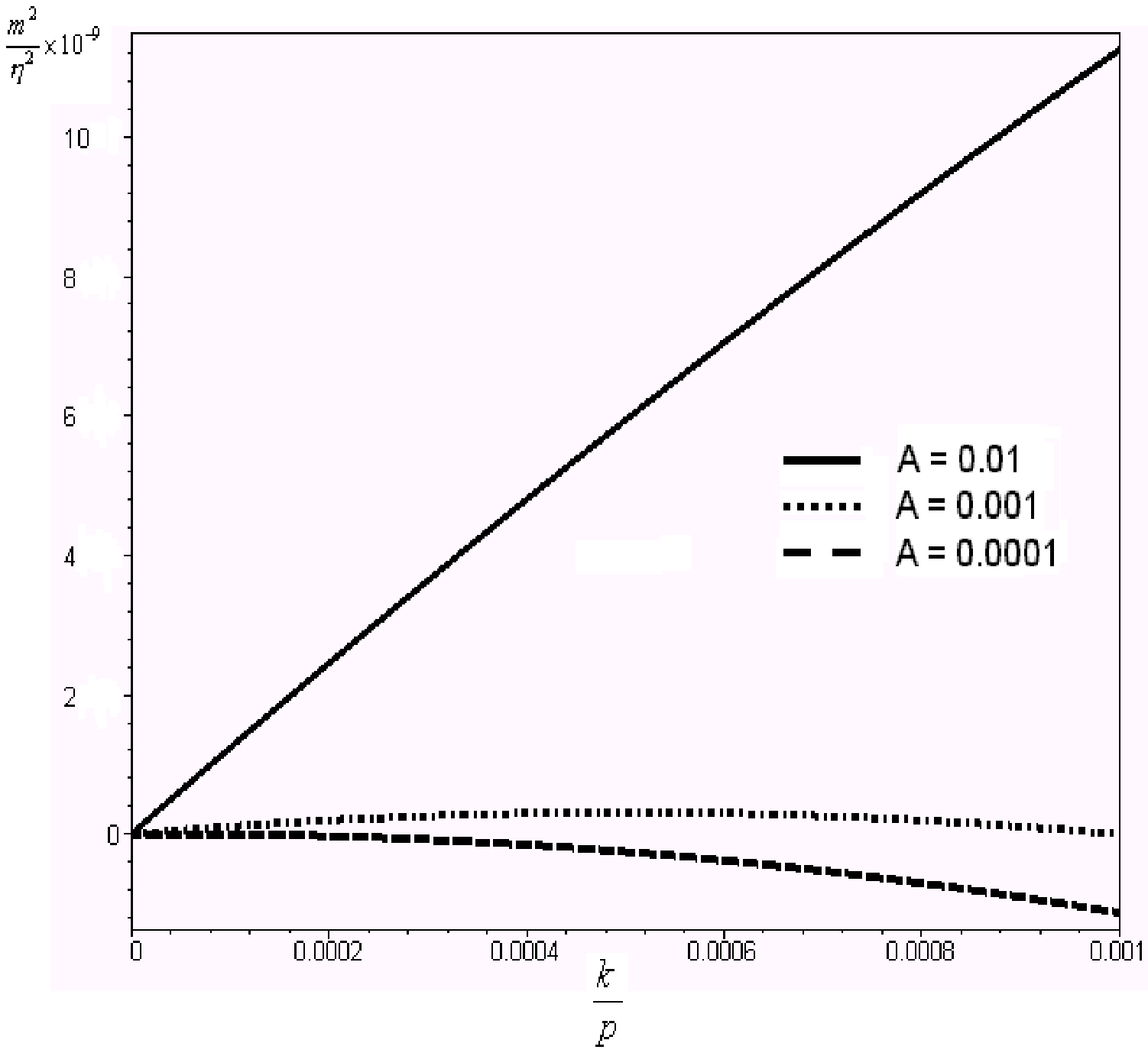, width=.9\textwidth}
\vskip -5 cm
\caption{ Plot of the ${m^2 \over \eta^2}$ in function of the ${k \over p}$ to the $0\leq {k\over p} \leq 10^{-3}$. }
\label{fig2}
\end{figure}

To analyse the possibility that  the parameters $m$ and $\eta $ give us a positive value, let us investigate the plot of  ${m^2 \over \eta^2}$ as a function of ${ k \over p}$, 
given in Fig (2). We see that there is a region where it is possible to obtain a range of values for which $m^2$ and $\eta^2 $ are both  positive.

Then, it is possible to find the stable potential to the cosmic string without flat directions.
The vacuum analysis of the potential gives us that it vanishes in its minimum.
So, SUSY is not spontaneously broken in the vacuum. The U(1) symmetry related with the vortex field $\phi_+$ is broken, but the extra $U(1)'$ symmetry related with the field $\sigma_+$ and $\sigma_-$ does not break and gives us an eletromagnetic propagation. In the core of the string,
this scenario changes: $<\phi_+> = <\phi_-> =0$ and $<\sigma_+> = \sqrt{{2\over e} (Ap + k)}$.
Then, the $U(1)'$ breaking in the core  gives us the bosonic particle condensate in the core. The potential does not vanish in these extrema, then the SUSY is broken in the core.

Until now, we have discussed only the spontaneous SUSY breaking. Now, let us explain 
in more details how the SUSY breaking appears in connection with the Lorentz breaking. To discuss this matter, let us recall that conventional Lorentz transformations are implemented as coordinate changes, and we usually refer to them as observer Lorentz transformations. However, we can also consider the so-called particle Lorentz transformations, which consist in applying boosts or rotations on particles and 
localised fields, but never on the background fields, contrary  to the observer Lorentz transformations, which act also on ackground fields.

Distinguishing between observer and particle Lorentz transformations is crucial for the kind of model we are considering here, where the  Chern-Simons term described in the Lagrangian of eq.(\ref{MCS}) is to be regarded as arising from a constant background field, $S^{\mu}= i\partial^{\mu}(S - S^*)$, which is to be seen as a global feature of the model, and is not related to localised experimental conditions, contrary the electromagnetic field, $A^{\mu}$, which is a perturbation that propagates in a space-time dominated by  $S^{\mu}$. We note that this part only exists if $S$ has an imaginary part.

So, in applying particle Lorentz transformations, the gauge invariant Chern-Simons term of eq.(\ref{MCS}) does not display Lorentz invariance, since $S^{\mu}$ is not acted upon by any $\Lambda $-matrix belonging to Lorentz group, so that the $\Lambda$'s acting on $\tilde F^{\mu \nu}$ and $A^{\alpha}$ do not combine to produce the $det \Lambda =1 $-factor  that would appear if $S^{\mu}$ were boosted, as it happens for the class of observer Lorentz transformations. In this particle frame, SUSY is broken. This detail shall be analysed in a forthcoming work,  where we analyse the case of an imaginary part of $S$ were $S$. In this letter, we consider only the real part of $S$. If this is the case (Re $S \neq 0$, Im $S =0$), then the only effect of the background is to renomalise the gauge field kinetic term and the Lorentz-violation Maxwell-Chern-Simons does not show up; in such a situation, no violation of Lorentz symmetry is detected.

\section{General conclusions}

In this work, we study a general supersymmetric action to present
a potential without flat directions. In our model, we study two aspects:
a gauge-field mixing and the case with a Maxwell Chern-Simons term.
We can compare these two possibilities and we see that in the Maxwell-Chern-Simons case, the potential has a minimum when the extra scalar field has a 
constant value. But,  we  pay attention to the fact that we only analyse the
real part of this field, since its complex part does not contribute to the analysis of the breaking. Nevertheless, if we consider the complex part, we have other
consequences to the model,such as the possibility of Lorentz symmetry breaking. If we analyse
the Lagrangian (\ref{lag1}) in components, we see that there appears a gauge-field mixing.
A motivation to the study of these potentials is a cosmic string scenario. This
approach is an alternative to the cosmic stirng potential without flat 
directions. In forthcoming  works, we must analyse the potentials in connection with 
cosmic string, bosonic supercondutivity and  Lorentz breaking
effects for  supersymmetric cosmic strings.

\end{document}